\documentclass[aip,jcp,twocolumn,reprint,citeautoscript,floatfix]{revtex4-1}
\usepackage{graphicx}
\usepackage{amssymb}
\usepackage{amsbsy}
\usepackage{multirow}
\usepackage{mathrsfs}
\usepackage{amsmath}
\usepackage[T1]{fontenc}
\DeclareMathAlphabet{\mathscrbf}{OMS}{mdugm}{b}{n}
\usepackage{bm}
\usepackage{color}
\usepackage{epstopdf}
\usepackage[normalem]{ulem}
\usepackage{verbatim}
\usepackage[table]{xcolor}

\usepackage{float}
\restylefloat{table}
\usepackage{placeins}
\usepackage{booktabs}
\usepackage{cleveref}

\usepackage{hyperref}

\usepackage[hyphenbreaks]{breakurl}

\usepackage{array,url,kantlipsum}

\newcommand{\drj}{\langle \Delta r_{\rm J}^2 \rangle}
\newcommand{\tw}{\langle t_{\rm w} \rangle}

\begin{document}
\author{Joseph F. Rudzinski}
\email{rudzinski@mpip-mainz.mpg.de}
\author{Marc Radu}
\author{Tristan Bereau}
\affiliation{Max Planck Institute for Polymer Research, Mainz 55128, Germany}

\title{
Automated detection of many-particle solvation states for accurate characterizations of diffusion kinetics
}

\begin{abstract}
Discrete-space kinetic models, i.e., Markov state models, have emerged as powerful tools for reducing the complexity of trajectories generated from molecular dynamics simulations.
These models require configuration-space representations that accurately characterize the relevant dynamics.
Well-established, low-dimensional order parameters for constructing this representation have led to widespread application of Markov state models to study conformational dynamics in biomolecular systems.
On the contrary, applications to characterize single-molecule diffusion processes have been scarce and typically employ system-specific, higher-dimensional order parameters to characterize the local solvation state of the molecule.
In this work, we propose an automated method for generating a coarse configuration-space representation, using generic features of solvation structure---the coordination numbers about each particle. 
To overcome the inherent noisy behavior of these low-dimensional observables, we treat the features as indicators of an underlying, latent Markov process.
The resulting hidden Markov models filter the trajectories of each feature into the most likely latent solvation state at each time step.
The filtered trajectories are then used to construct a configuration-space discretization, which accurately describes the diffusion kinetics.
The method is validated on a standard model for glassy liquids, where particle jumps between local cages determine the diffusion properties of the system.
Not only do the resulting models provide quantitatively accurate characterizations of the diffusion constant, but they also reveal a mechanistic description of diffusive jumps, quantifying the heterogeneity of local diffusion.
\end{abstract}
\maketitle



\newpage
\section{Introduction}
Molecular dynamics (MD) simulations of condensed-phase systems provide a wealth of microscopically-detailed information about complex dynamical processes.
The extraction of useful insight from MD trajectories relies on simplifying the underlying, high-dimensional free-energy landscape to a handful of essential degrees of freedom that retain an accurate description of the processes of interest.
In recent years, Markov state models (MSMs) have emerged as powerful tools for reducing the complexity of MD data through a discrete-time and -space description of the system's dynamics.\cite{Bowman:2014}
Indeed, these simple kinetic models have been successfully employed to describe complex processes in biomolecular systems, e.g., protein folding,~\cite{Chodera:2011,Lane:2011,Zhuang:2011} binding,~\cite{Buch:2011,Plattner:2015,Shukla:2016} and allostery.~\cite{Bowman:2012}
For these systems, there are a set of well-established, low-dimensional order parameters, e.g., the torsional angles along the peptide backbone, which provide an excellent starting point for characterizing the conformational landscape.
Using standard dimensionality reduction techniques,~\cite{Altis:2008, Schwantes:2015} these order parameters may be combined with other potentially important degrees of freedom to generate collective variables on which accurate and robust MSMs can be built.

There are significantly fewer examples of MSMs applied to describe diffusion in condensed-phase systems.
Traditionally, Stillinger and Weber's inherent structure theory for liquids provides a general starting point for configuration-space discretizations in many-particle systems.~\cite{Stillinger:2014}
Despite successes in characterizing the potential energy landscapes for a broad range of systems,\cite{Wales:2015} the approach becomes prohibitively expensive as the system size and complexity increases.
Alternative approaches bypass the calculation of inherent structures by employing an explicit discretization along the global coordinate system,~\cite{Teo:2013,Pintus:2015,Cardenas:2016} although the coordinate dependence may lead to difficulties in aligning models built from distinct simulation trajectories.

The problem can be simplified significantly by considering the diffusion properties of individual molecules and characterizing configuration space with respect to their local environment.
For example, Rao and coworkers~\cite{Rao:2010,Prada-Gracia:2012} have demonstrated that the hydrogen-bonding dynamics in liquid water can be well described by an MSM, employing a graph representation of the hydrogen bonding network to represent the relevant set of solvation states.
The transferability of this approach to distinct or less-ordered liquids has not yet been investigated.
In a different study, Sodt {\it et. al.}~\cite{Sodt:2014} used more generic properties of the local composition to build a hidden Markov model for describing ordering dynamics in a binary lipid mixture, probing the relationship between low-order descriptors of solvation and the ``true'' set of underlying solvation states.

A more general investigation into the relevant order parameters for describing activated diffusion processes has been performed in the field of glassy liquids.~\cite{Ciamarra:2016}
While several studies have employed inherent structure potential energies as an order parameter for characterizing particle jumps in these systems,~\cite{Doliwa:2003,Li:2011,Helfferich:2016} other researchers have attempted to find simpler descriptions of these jumps.~\cite{Vollmayr-Lee:2004,Helfferich:2014}
For example, de Souza and Wales~\cite{deSouza:2008} investigated coordination-number-based order parameters for characterizing the cage-breaking process in a binary Lennard-Jones mixture.
They demonstrated that these order parameters suffer from being rather noisy indicator functions of a cage break and, consequently, may be incapable of resolving reversal events, significantly degrading the resulting description of diffusion.
Moreover, their analysis employed a user-specified definition of cage breaking, complicating applications to other systems.

In this work, we propose a systematic method for constructing configuration-space representations for diffusion processes in condensed-phase systems, without relying on previous physical or chemical intuition about the system.
Instead of working with higher-dimensional properties of the solvation state (e.g., hand-crafted definitions of a cage break), we restrict ourselves to features that are simple functions of the instantaneous coordination number (CN) about each particle.
The methodology can also be extended to features that describe orientational or 3-body correlations, which may be important for characterizing more complex liquid solvation structures.~\cite{Steinhardt:1983,Pluharova:2017,Gasparotto:2016,Giberti:2017}
Following the work of Sodt {\it et. al.}~\cite{Sodt:2014}, we treat the CN features as indicators of an underlying, latent Markov process, in order to overcome the inherent noisy behavior of CN-based observables.
From time series of each coordination number feature, we first infer a hidden Markov model describing the dynamics of the hidden variables through probabilistic emissions to the observed data.
We then employ this model to transform the observations into trajectories of the underlying (i.e., hidden) states of the corresponding solvation feature.
We subsequently perform a dimensionality reduction on these filtered trajectories to extract a low-dimensional free-energy landscape, upon which a coarse configuration-space representation is built.
From the resulting metastable solvation states, a continuous time random walk (CTRW) model of the underlying diffusion process is built, revealing a transparent picture of single-particle diffusion.

We consider single-particle dynamics in a standard model for glassy liquids, and demonstrate that the proposed method results in a structured, low-dimensional free-energy landscape that accurately describes the relevant diffusion processes, i.e., cage breaks.
In particular, we verify the accuracy of the automated configuration-space representation by demonstrating that the diffusion constant, as determined from the CTRW model, is quantitatively accurate.
The general utility of the approach is clearly demonstrated through the mechanistic picture of diffusive jumps and, in particular, the characterization of local dynamic heterogeneity, which arises from an analysis of the resulting kinetic networks.

\section{Methods}

\subsection{Model and simulations}

\begin{figure}[ht]
\centering
\includegraphics[width=\linewidth]{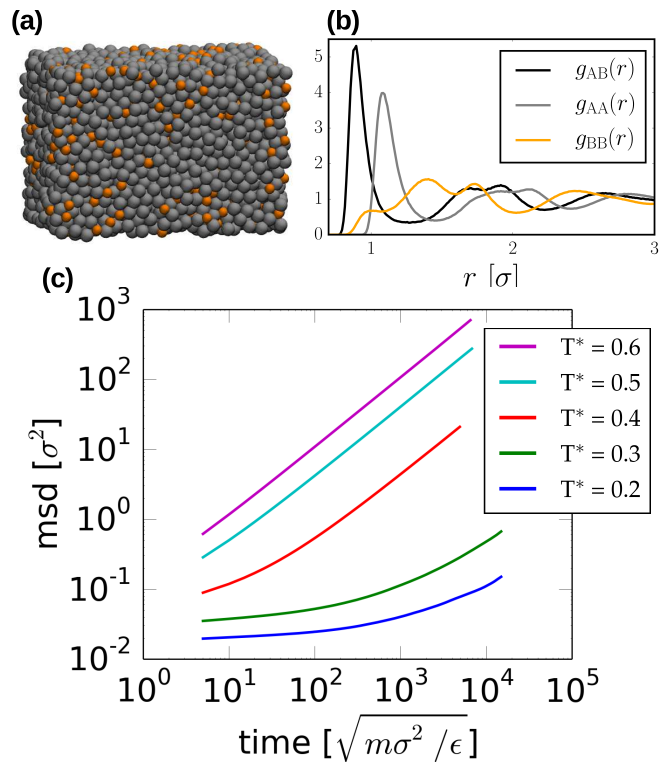}
\caption{
Characterization of the simulation models.
(a) Representative snapshot demonstrating the mixing of A (gray) and B (orange) particles.
(b) Representative set of radial distribution functions from the $T^* = $~0.4 simulations.
(c) Mean square displacement (msd) of particle positions as a function of time obtained from simulations over a range of temperatures.
}
\end{figure}

In this work, we consider the Kob-Andersen binary Lennard-Jones model\cite{Kob:1995} as a representative  system for diffusion in complex liquids.
This model has two particle types, A and B, interacting according to Lennard-Jones potentials with $\epsilon_{\rm{AA}}$ = 1, $\sigma_{\rm{AA}}$ = 1, $\epsilon_{\rm{AB}}$ = 1.5, $\sigma_{\rm{AB}}$ = 0.8, $\epsilon_{\rm{BB}}$ = 0.5, and $\sigma_{\rm{BB}}$ = 0.88.
Both particle types have the same mass: $m_{\rm A} = m_{\rm B} = m = 1$.
We considered a system of 4000 A particles and 1000 B particles in a rectangular box with $L_{\rm x} \approx 1.5L_{\rm y} = 1.5L_{\rm z}$.
All simulations were performed using the Espresso++ simulation package~\cite{Halverson:2013} in the $NPT$ ensemble at a constant pressure of 0 $\epsilon/\sigma^3$, where $\sigma \equiv \sigma_{\rm{AA}}$, while employing the Berendsen barostat~\cite{Berendsen:1984} and a Langevin thermostat.~\cite{Huenenberger:2005} 
A timestep of 0.005~$t^*$ was used, where $t^* = t\sqrt{m \sigma^2 / \epsilon}$ is the reduced time.
Initially, we considered five different reduced temperatures: $T^* = k_{\rm B}T/\epsilon = \{0.6, 0.5, 0.4, 0.3, 0.2\}$.
After equilibration, each system was simulated for a total of 25000~$t^*$, saving configurations every 0.5~$t^*$, for a total of 50,000 configurations per system.
From these simulations, the average reduced densities were determined to be $\rho^* = \rho \sigma^3 = \{1.04, 1.09, 1.13, 1.17, 1.19\}$, respectively.

Fig.~1 presents structural and dynamical characteristics of the model.
Panel $(a)$ illustrates the structure of the liquid with a simulation snapshot, while panel $(b)$ presents a representative set of radial distribution functions between the various pair types from the $T^*$ = 0.4 simulation.
Panel $(c)$ presents the mean squared displacement of the particle positions as a function of time for each simulation temperature.
This plot demonstrates that as the temperature is lowered, the subdiffusive region of the mean square displacement quickly extends over a very large range of time scales, indicating glassy dynamics and making it difficult to accurately sample configuration space.~\cite{Barrat:1991}
As a consequence of the latter, we restrict ourselves to the higher three temperatures in this study, where the long-timescale dynamics can be accurately determined directly from the simulation data.
In this way, we can assess the accuracy of our coarse-grained, kinetic description of dynamics, as described further below.

\subsection{Weighted coordination numbers (WCNs)}
Our aim is to build a low-dimensional representation of configuration space capable of describing single-particle diffusion.
This requires the determination of order parameters that accurately characterize the local environment of each particle.
Ideally, these order parameters should be constructed in an automated fashion from a set of general descriptors of solvation.
The instantaneous coordination number (CN) of a particle (i.e., the number of particles in a surrounding spherical shell at a particular time), is perhaps the simplest example of such a descriptor.
CN-based features suffer from at least two major limitations: $(i)$ the use of strict cut-off values to indicate whether a surrounding particle is counted as in or out of the solvation shell results in discontinuous time series and $(ii)$ their low-dimensionality provide noisy and incomplete characterizations of the solvation state.

To ease the former problem, we employ ``weighted coordination numbers'' (WCNs), which use a Gaussian function to weight the contribution of each surrounding particle to individual solvation features based on the outer particle's distance from the center particle.
We first identify relevant solvation features as the various maxima along each radial distribution function.
We then place a Gaussian weighting function at the center of each of these features (Fig.~2).
These Gaussians are normalized to 1 at their maximum.
The width of each Gaussian was chosen based on the position of the neighboring Gaussians, such that the value of the intersection point of the Gaussians was either 0 or approximately 0.25, depending on if the solvation features were largely separated or highly overlapping, respectively.
However, we do not expect the results to be particularly sensitive to the chosen overlap.
Then, the WCN of each particle for a particular surrounding particle type (A or B) and a particular feature (given the pair type, e.g., 1-4 for A-B pairs) was determined as a function of time by summing up the weights for each configuration. 

Panel $(a)$ of Fig.~3 presents an illustrative example of such a trajectory, for B particles surrounding a specific A particle, using the weights from the first solvation feature for A-B pairs, $w^{(1)}_{\rm AB}(r)$.
This time series characterizes the number of particles residing close to the first solvation distance around this particular A particle.
The WCN property counts a particle sitting directly at the center of this feature as a whole particle, while particles further away from the feature center are counted as a fraction of a particle, as dictated by the Gaussian weighting function.
In this way, there are smoother transitions between solvation features.

\begin{figure}[ht]
\centering
\includegraphics[width=\linewidth]{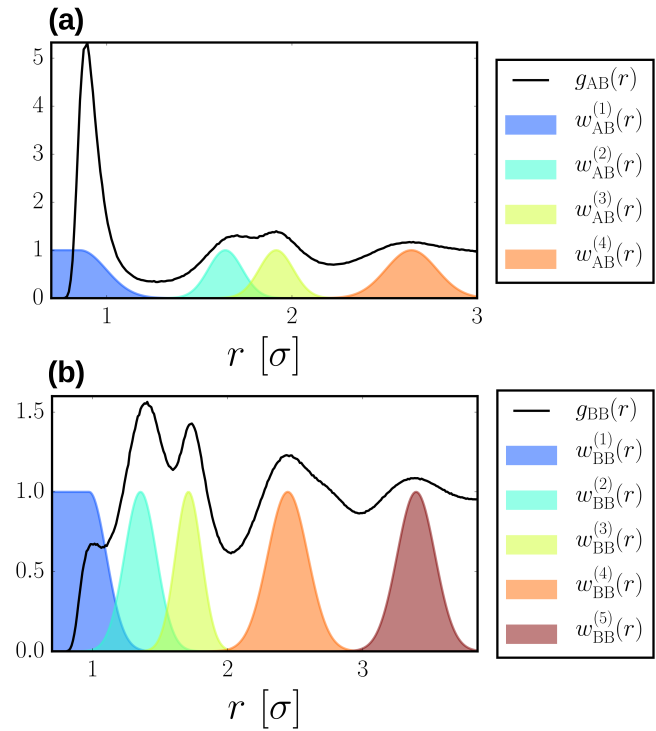}
\caption{
Gaussian weights (colored distributions) for determining the weighted coordination number, as described in the main text.
Radial distribution functions (solid black curves) are presented between A and B pairs (a) and between pairs of B particles (b).
}
\end{figure}

\subsection{Hidden-Markov-model filtering}

\begin{figure*}[bt!]
\centering
\includegraphics[width=\linewidth]{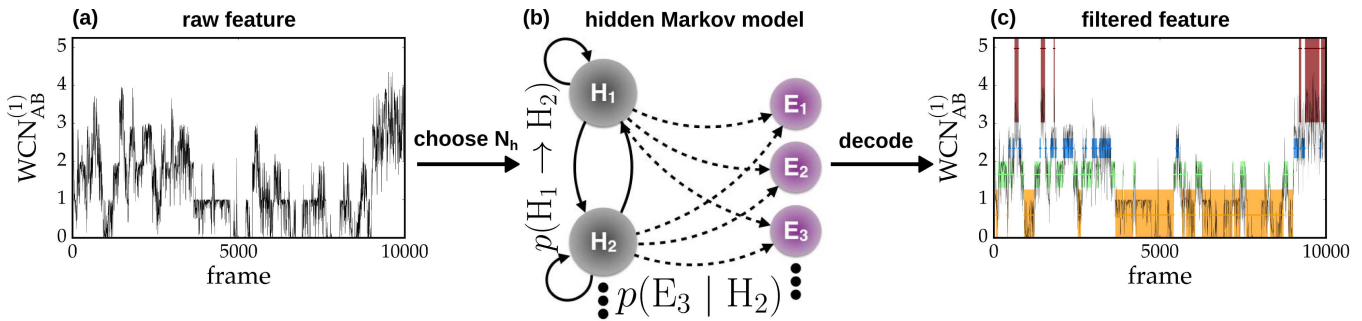}
\caption{
Workflow for hidden-Markov-model filtering.
First, for each particle, trajectories of the weighted coordination numbers (a) are determined using the weights presented in Fig.~2, for each pair type and for each feature in the corresponding radial distribution function.
Then, a hidden Markov model with a specified number of hidden states is parametrized from all the simulation trajectories of a given feature (b).
Here, $p({\rm H_1} \rightarrow {\rm H_2})$ represents the transition probability from hidden state $\rm H_1$ to hidden state $\rm H_2$, while $p({\rm E_3} \mid {\rm H_2})$ represents the probability of observing emission state $\rm E_3$ given that the system is in hidden state $\rm H_2$.
Using this model, each trajectory is filtered by determining the most likely hidden state (transparent blocks) at each time step (c).
}
\end{figure*}

The WCNs contain information about the solvation state of each particle, but are imperfect indicators of this underlying state.
For this reason, we treat the solvation state as a latent variable and construct a hidden Markov model (HMM),~\cite{Rabiner:1986} based on the information in the WCNs, to predict the underlying state.
HMMs describe a Markov process occurring between a set of $n$ hidden states, through their probabilistic relationship to a set of $N$ emission signals or observed states.
HMMs are often employed as non-linear filters that infer the most probable hidden state of the system, given a set of observations.
The parameters of this model correspond to the set of $n^2$ transition probabilities between pairs of hidden states as well as the set of $nN$ emission probabilities---the probability that a given hidden state gives rise to the observation of a particular emission state.
Panel $(b)$ of Fig.~3 presents a schematic of the HMM structure.

Ideally, all WCN information should be considered simultaneously to determine the HMM.
However, this results in an HMM with many parameters which may be difficult to parametrize.
Instead, for each pair type and solvation feature, we parametrized a separate HMM from the WCN trajectories of the corresponding particles using the Baum-Welch algorithm.~\cite{BaumWelch}
We employed the trajectories of 1000 particles for parametrization of the HMM, allowing cross validation for the A particles (see the ``Cross validation of HMMs'' section in the Supporting Information for details).
For each HMM, the emission states correspond to a discretization of the respective WCN distribution into 20 bins, although we do not expect the results to be particularly sensitive to the exact number of chosen emission states.
Based on a Markovianity criterion for the resulting HMMs, we constructed all HMMs with 4 hidden states using a lag time of 4~$t^*$ (see the ``Selection of hyperparameters'' section of the Supporting Information for details).

Each HMM characterizes the relationship between the (hidden) solvation state of the particle and the observed WCN value, and can be used to filter the WCN time series into trajectories of the most likely hidden state, as illustrated in panel $(c)$ of Fig.~3.
The Viterbi algorithm~\cite{Viterbi} was employed to perform this ``decoding'' of the trajectories, applied at the level of resolution of the HMM (i.e., trajectories were pruned with a spacing of 4~$t^*$ for filtering).

\subsection{Dimensionality reduction and clustering}

The filtered WCN trajectories describe the hidden state of each particle as a function of time with respect to the type of surrounding particle and particular solvation feature.
However, because the HMMs for each feature were constructed independently, these filtered trajectories will not align perfectly in general.
As a consequence, we perform a dimensionality reduction to detect correlations in the hidden states predicted using HMMs for distinct features.
Since the HMM filtering procedure already performs a nonlinear transformation from the emission to hidden states of a particle, we apply a linear dimensionality reduction scheme, rather than considering higher-dimensional schemes.~\cite{Ceriotti:2011,Rohrdanz:2011,Sultan:2018,Jong:2018}
Principal component analysis (PCA) was applied to the complete set of filtered trajectories for each particle type, while treating the particles as indistinguishable, to construct a low-dimensional configuration-space representation of the local solvation state of a single particle.
In each case, only the two ``most significant'' PCA dimensions were retained for clustering, since the trajectory data was approximately Gaussian distributed along subsequent dimensions.

Fig.~4 presents a representative free-energy surface along the two principal components obtained using the raw (a) and filtered (b) set of trajectories.
The emergence of clear conformational basins in panel $(b)$ demonstrates significant correlations between the predicted hidden states along individual WCN trajectories.
As discussed in more detail in the ``Solvation state characterization'' section below, the basins with positive PC$_1$ values correspond to solvation states depleted with respect to surrounding B particles, while those with negative values correspond to solvation states with an increased number of surrounding B particles relative to the overall average.
The Supporting Information section presents a more detailed analysis of the individual basins.
In contrast to panel $(b)$, panel $(a)$ demonstrates that the noise in the raw WCN trajectories effectively wipes out any correlations between the various features.
Thus, Fig.~4 demonstrates the role of the filtering procedure in generating a useful configuration-space representation that characterizes distinct features of the solvation environment.

\begin{figure}[h!]
\centering
\includegraphics[width=\linewidth]{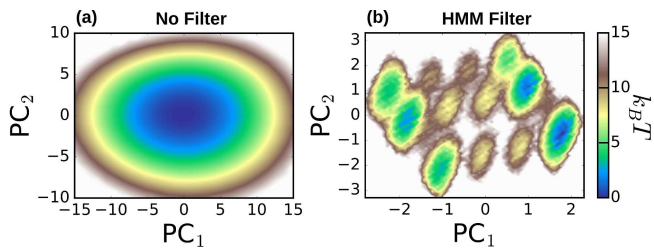}
\caption{
Using the unfiltered (a) and filtered (b) WCN trajectories as input, principle component analysis is applied to obtain a low-dimensional representation of configuration space.
}
\end{figure}


We note that while time-lagged independent component analysis is often employed for dimensionality reduction for building kinetic models,~\cite{Schwantes:2015} this procedure yielded an insufficient description of solvation states.
This may have to do with the uncertainty of the filtering procedure in transition regions of the trajectory.
However, any resulting dynamical artifacts are effectively removed by the coarse-graining and coring procedure described below (see the ``Cross validation of HMMs'' section of the Supporting Information for a more detailed discussion of these potential artifacts).

With a low-dimensional description of solvation states in hand, we aimed to determine a small number of relevant metastable solvation states to describe the underlying diffusion process.
First, we generated 50 microstates which spanned the 2-D PCA space (see Fig.~S18), using the $k$-means clustering algorithm.~\cite{Lloyd:1982}
We then constructed a Markov state model~\cite{Bowman:2014}---a discrete-time and -space kinetic model---in terms of these microstates, using the {\sc pyEmma} software package.~\cite{Scherer:2015}
Based on a standard implied timescales test,~\cite{Bowman:2014} we employed lag times of approximately 25~$t^*$ for these models.
We then applied the Perron cluster analysis technique (PCCA+)~\cite{Deuflhard:2005}---a method for systematic coarse-graining of a Markov state model---to reduce the 50 microstate representation to between 2 and 6 metastable states.
We considered an increasing number of metastable states, beginning with 2, until the uncertainty of the kinetic model reached a predefined threshold.
PCCA+ outputs a set of membership probabilities, i.e., the probability that a microstate belongs to a particular metastable state.
We constructed the metastable states by requiring that microstates belonging to a particular metastable state have a membership probability of greater than 0.75.
As a consequence, the metastable states do not represent a strict partition of configuration space.
We then generated trajectories along the metastable solvation states by specifying that a particle remains within its current metastable state until a new metastable state is reached.
This simple ``coring'' procedure effectively reduces recrossing artifacts due to an imperfect description of the dividing surfaces between metastable states, as described elsewhere.~\cite{Jain:2012}
From these cored trajectories, we determined the distribution of $(i)$ waiting times in metastable state $i$, before transitioning to metastable state $j$, $\{t_{\rm{w}}\}_{ij}$ and $(ii)$ squared jump lengths for transitions from state $i$ to state $j$, $\{\Delta r^2_{\rm{J}}\}_{ij}$, using the particle coordinates from the original trajectories (see Fig.~S25 for representative distributions). 


\section{Results and Discussion}

We have proposed a methodology for constructing a low-dimensional configuration-space representation from simple and general descriptors of local solvation, using a hidden-Markov-model (HMM) filtering technique.
We applied the methodology to the Kob-Andersen binary Lennard-Jones mixture at three different temperatures, resulting in 2- to 6-state representations for each system. 

\subsection{Solvation state characterization}

To understand the identity of the metastable solvation states, we determined the distribution of weighted coordination numbers (WCNs) sampled while the system resides in each particular state.
We found that these distributions are often highly overlapping, with an increased overlap at higher temperatures.
Nonetheless, the differences between these distributions are sufficient for distinguishing distinct solvation states.

\begin{figure}[ht!]
\centering
\includegraphics[width=\linewidth]{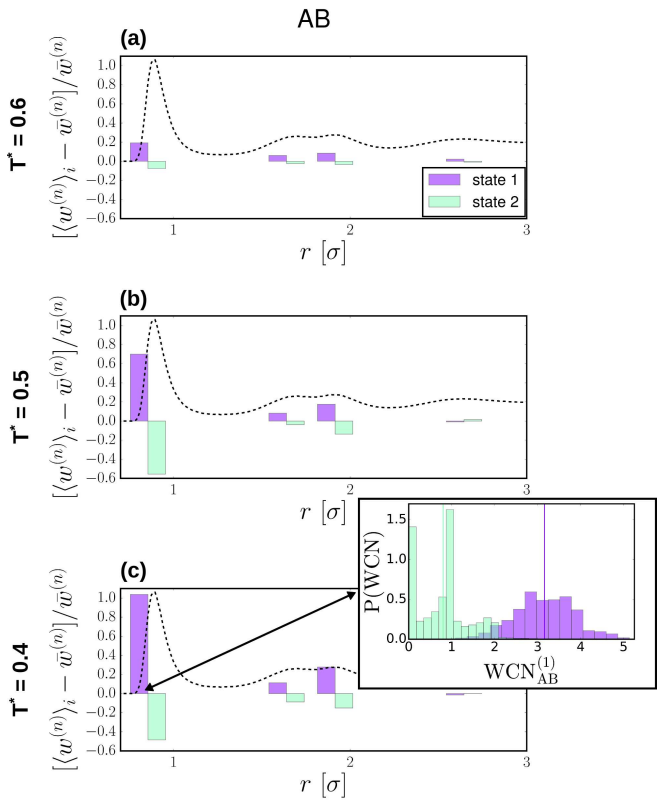}
\caption{
Characterization of a 2-state representation of solvation for A particles.
The transparent bars quantify the difference in the average WCN within a metastable state relative to the ensemble average.
The radial distribution functions (dashed black curves), included as reference for the feature definitions, are rescaled to the magnitude of the presented data.
The inset in the bottom panel presents the distributions of the first solvation shell WCN for the two metastable states.
}
\end{figure}

Fig.~5 quantifies the difference between the distributions of each WCN feature, $n$, for A particles in terms of surrounding B particles.
This difference is measured by the deviation of the average value of the WCN distribution for a particular metastable state $i$, $\langle w^{(n)} \rangle_i$, relative to the overall average, $\bar{w}$: $[\langle w^{(n)} \rangle_i - \bar{w}^{(n)}]/\bar{w}^{(n)}$.
Note that here $w$ refers to the total value of the WCN and not an individual Gaussian weight. 
For simplicity, Fig.~5 presents results for the 2-state models (see Figs.~S22-S24 for characterization of the multi-state models).
We present WCN features from the A-B radial distribution, since these distributions are more informative for the definition of solvation states than the A-A radial distribution features.
The 2-state model characterizes solvation in terms of a state solvated with B-particles (state 1) and a state depleted of B particles (state 2), relative to the average.
These states are anti-correlated with the distribution of A particles (Fig.~S20).
The three panels in Fig.~5 demonstrate a consistent metastable state definition over the various temperatures, although the deviation from the average WCN values for each state diminish as temperature increases.
The inset in the bottom panel illustrates the (mild) overlap of the WCN distributions for the first solvation shell feature.

\begin{figure}[ht!]
\centering
\includegraphics[width=\linewidth]{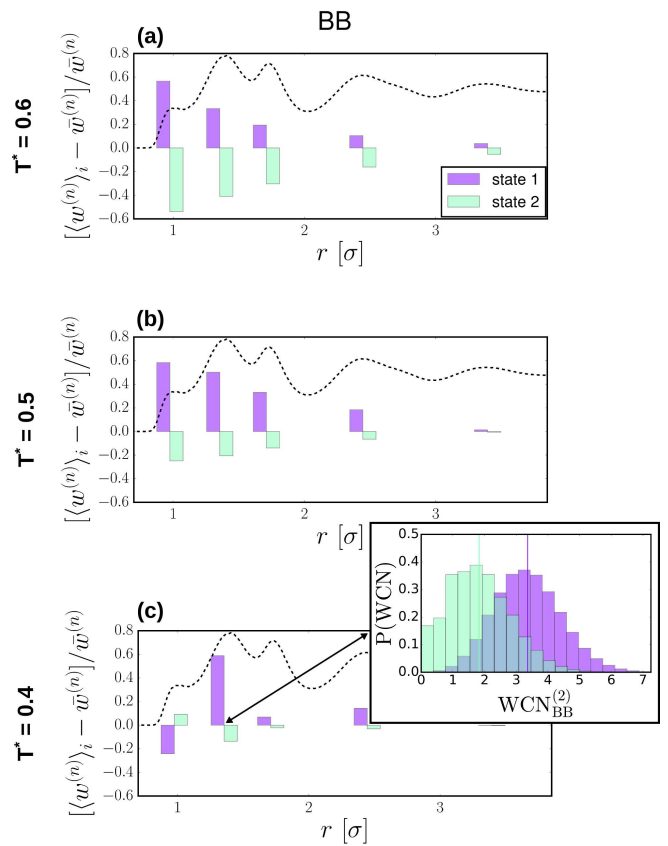}
\caption{
Characterization of a 2-state representation of solvation for B particles.
The transparent bars quantify the difference in the average WCN within a metastable state relative to the ensemble average.
The radial distribution functions (dashed black curves), included as reference for the feature definitions, are rescaled to the magnitude of the presented data.
The inset in the bottom panel presents the distributions of the first solvation shell WCN for the two metastable states.
}
\end{figure}

Similar to Fig.~5, Fig.~6 characterizes the difference in the WCN distributions for the B-B radial distribution features of a 2-state representation.
The distribution of B particles around the center particle again largely dictates the solvation state.
Similar to the solvation of A particles, the 2-state representation characterizes solvated (state 1) and depleted (state 2) states in terms of surrounding B particles.
For the lowest temperature, these states correspond to the free-energy basins with positive and negative PC$_1$ values, respectively, in Fig.~4b.
The bottom panel of Fig.~6 demonstrates an inconsistency in the definitions of the metastable states in terms of the first feature of the radial distribution function at the lowest temperature.
This discrepancy is probably due to poor sampling in the low temperature simulation, since B-B pairs only seldomly come within the distance corresponding to the first feature.
Based on the PCA, we determined that this shoulder feature plays only a minor role in determining the overall solvation state of the particle.
As a result, this apparent artifact does not affect the accurate description of the diffusion process, as demonstrated below.

\subsection{Properties of single-particle diffusion}

\subsubsection{Particle jumps in structural glasses}

The single-particle dynamics of supercooled liquids and structural glasses have been extensively studied, in an attempt to understand the disparate behavior of structural versus dynamical properties upon cooling.
In these systems, particles diffuse via a cage-breaking mechanism, where individual particles are trapped for relatively long periods of time oscillating in local cages, before spontaneously undergoing larger-scale transitions until relaxing into a new cage.~\cite{Ciamarra:2016}
Many theories have been proposed to describe these jumps using, e.g., mode-coupling, random first order, and free volume approaches,~\cite{Goetze:2008} although the precise connection between these descriptions remains elusive.~\cite{Ciamarra:2016}
One of the simplest approaches to the relaxation dynamics in structural glasses is the adoption of a continuous-time random walk (CTRW) description,~\cite{Montroll:1965} which assumes the absence of spatial and temporal correlations between jumps.
Although correlations are known to occur in reality, even at high temperatures, this approximation is of use for simplifying the description of jumps for clearer insight into the diffusion mechanism.
In practice, one must first choose the precise definition of a jump before determining properties of the diffusion dynamics such as the average squared jump length, $\drj$, and waiting time between jumps, $\tw$.
Since jumps are assumed to be uncorrelated, the diffusion constant can be determined in terms of these average properties:~\cite{Doliwa:2003}

\begin{equation}
D = \frac{\drj}{6 \tw}
\label{eq-D}
\end{equation}

De Souza and Wales~\cite{deSouza:2008} investigated the CTRW description in the context of a binary Lennard-Jones mixture.
They investigated to what extent low-dimensional features of the solvation structure about a particle could be used to define a cage break.
In particular, they tracked the number of particles within a cutoff distance of a center particle and defined jumps as occurring when two particles simultaneously left the solvation shell.
From this definition, they built a CTRW model and demonstrated that this representation provided a relatively accurate characterization of the diffusion constant, but only when they adjusted the jump definition to account for reversal events, where particles returned to their cage after the initial break.
Although quantitatively accurate at high densities and low temperatures, their CTRW models demonstrated significant errors at lower densities and higher temperatures, supposedly due to an insufficient effective dividing surface dictated by their jump definition.

\subsubsection{CTRW models for multi-state representations of particle jumps}

In this work, we examine a CTRW representation of particle jumps in the Kob-Andersen model.
As described in more detail in the methods section, we employ instantaneous weighted coordination numbers---measures of the number of particles within spherical slices about a central particle---as indicators of the solvation state of a particle.
We employ a hidden-Markov-model filter to predict the underlying state of solvation from the noisy trajectories provided by these features.
Using the filtered trajectories, we determine a coarse representation of the local solvation state (2-6 states), and then construct a CTRW model to describe single-particle diffusion in the underlying simulation.

We extend the CTRW formalism to our multi-state representation of diffusion by determining the ``local'' diffusion constants between pairs of states in our model:

\begin{equation}
D_{ij} = \frac{\drj_{i \rightarrow j}}{6 \tw_{i \rightarrow j} } ,
\end{equation}
where $i$ and $j$ represent metastable solvation states in our representation, and the ensemble averages are determined from the filtered trajectories (using coordinate data from the original simulation trajectories).
We set diagonal terms of this local diffusion matrix to zero by definition.
The total diffusion can then be determined as a weighted sum of local diffusion constants:

\begin{equation}
D = \sum_{ij} \pi_i D_{\rm ij} ,
\end{equation}
where $\pi_i$ is the equilibrium probability of state $i$.

Fig.~7 presents the diffusion constants calculated from the CTRW models for various multi-state representations, compared with the exact values determined in the conventional way from the original simulation trajectories.
The error bars were calculated by determining the standard error from the jump length and waiting time distributions and then propagating these errors into the diffusion constant assuming no correlation between the variables.
Remarkably, the CTRW models determine nearly quantitatively accurate diffusion constants, with exception of the low temperature models for A particles, which demonstrate small deviations.

\begin{figure}[ht!]
\centering
\includegraphics[width=\linewidth]{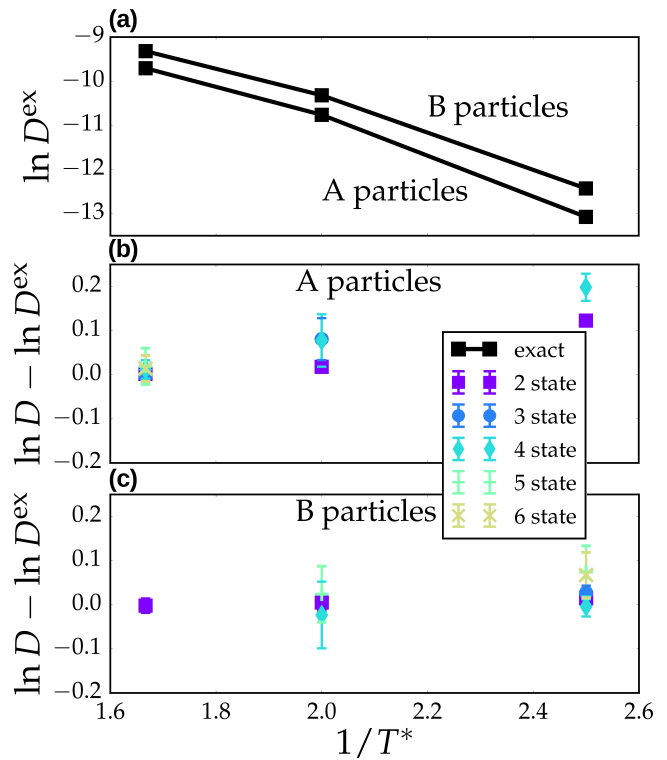}
\caption{
Diffusion constants for A and B particles determined from the original simulation trajectories (``exact'', black square markers) and also from the CTRW models with varying number of metastable states (colored with various markers).
}
\end{figure}

However, the various multi-state representations are not possible in every case.
For example, at high temperature, only a 2-state representation is possible for describing the diffusion of B particles.
Representations with a higher number of metastable states resulted in rarely visited solvation states and, thus, insufficient statistics for the waiting time and jump length distributions.
As the temperature increases, thermal noise blurs the identity of distinct solvation states, making multi-state representations of solvation difficult.
This issue is analogous to the errors observed in previous work at higher temperatures with hand-crafted definitions of diffusion jumps.~\cite{deSouza:2008}
The onset of these problems for B particles and not A particles at $T^* = 0.6$ likely has to do with the larger size of the A particles, resulting in better separated features in the A-A and A-B radial distribution functions relative to B-B pairs.
Fig.~7 also demonstrates a limited resolution for A particles at low temperatures, although this is more likely due to issues of statistics (as explained in more detail in the ``Cross validation of HMMs'' section in the Supporting Information) 

Direct comparison to the investigation of de Souza and Wales~\cite{deSouza:2008} is not possible, due to differences in model parameters and simulation set up.
Our results strongly suggest that the automated configuration-space representations are at least as accurate as employing manual definitions.
This is significant, since the manual definitions utilized specific knowledge about the system dynamics and explicitly accounted for reversal events during the particle jumps.
Further investigation is necessary to assess whether the automated scheme can outperform manual definitions in certain regimes.
However, the benefit of our proposed scheme is clear, since it employs generic order parameters, in lieu of system-specific features or insight.

\subsubsection{Heterogeneity of local diffusion}

The potential of the proposed methodology extends beyond the description of long-time diffusion with a CTRW model.
The multi-state representations of solvation provide a platform for investigating the mechanisms of single-particle diffusion.
For the Kob-Andersen model, the known cage-breaking mechanism is consistent with the hopping between local solvation states in our models, although the perspective is slightly different.
Normally, jumps between local cages are described without specifying the identity of the local cage.
(Note that here we are effectively ignoring transitions between identical cages).
With these extra details, we can investigate local features of particle jumps, e.g., the difference in local diffusion behavior between particular pairs of metastable states.

\begin{figure}[t!]
\centering
\includegraphics[width=\linewidth]{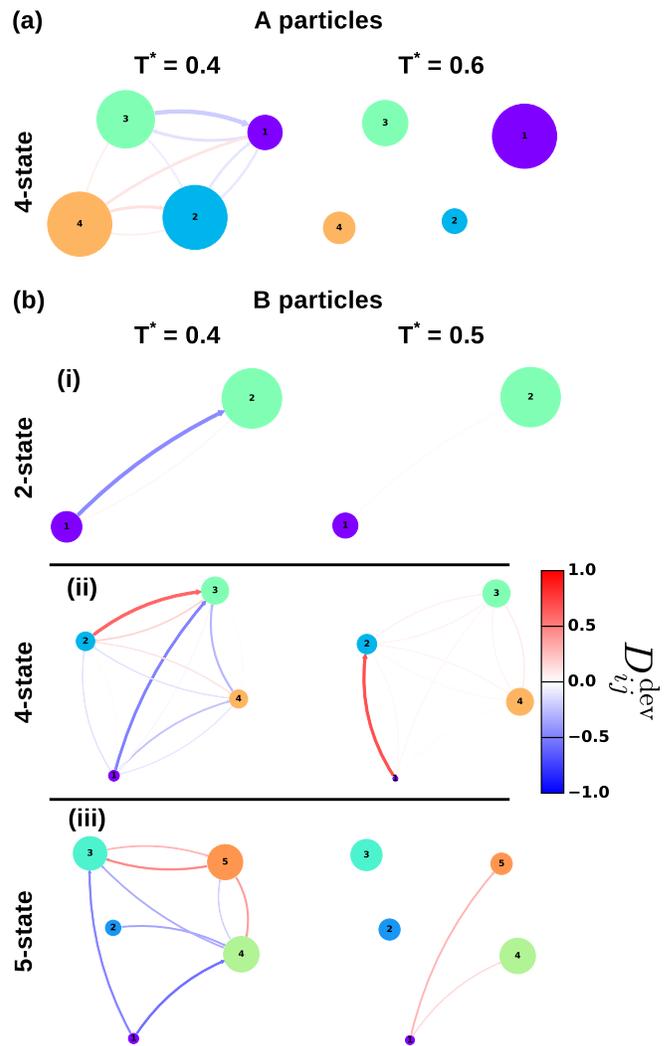}
\caption{
Characterization of local diffusion networks.
Network representations of the multi-state CTRW models, plotted along the 2-D PCA coordinates obtained from the $T^* = 0.4$ simulation.
The node sizes represent the relative population of each metastable state.
The arrows illustrate the deviation, $D^{\rm{dev}}_{ij} = \pi_i \frac{(D_{ij} - D)}{D}$, of the local diffusion constants (between pairs of states $i$ and $j$) from the total average.
}
\end{figure}

Fig.~8 presents network representations of various multi-state models describing the single-particle diffusion of A (a) and B (b) particles.
We consider models with various numbers of chosen states to ensure that our conclusions are robust to the precise network representation.
The network is plotted in the 2-D PCA space of the lower temperature system.
The node positions correspond to the average position when particles are in the corresponding metastable state at this temperature.
It is important to note that, because the network states were determined independently for each particle type and simulation temperature, there is no direct correspondence between metastable states at different temperatures.
However, by analyzing the characteristics of the metastable states, as in Figs.~5 and 6, we can make approximate comparisons between states.
The size of the node represents the relative population of each metastable state.

Using the definition of the local diffusion constant, $D_{ij}$, we determined the deviation from the total diffusion constant, $D^{\rm{dev}}_{ij} = \pi_i \frac{(D_{ij} - D)}{D}$, which characterizes the heterogeneity of local diffusion in the solvation network.
The relative deviation is weighted by the stationary distribution of state $i$, to quantify the relative contribution to the total diffusion constant.
That is, it is possible that a transition has a very large deviation from the average diffusion but is so rare that it does not significantly contribute to the diffusion constant.
Such transitions will be effectively ignored by this metric.

The network arrows in Fig.~8 represent the various $D^{\rm{dev}}_{ij}$ values between pairs of metastable states.
Red (blue) arrows denote local diffusion constants that are larger (smaller) than the overall average.
Transitions with $D^{\rm{dev}}_{ij} < 10^{-6}$ were removed for clarity.
Fig.~8 demonstrates that as the temperature decreases, the local diffusivities become increasingly heterogeneous for both A and B particles.
This observation is consistent with the emergence of dynamical heterogeneities as the system enters the glassy regime.~\cite{Ciamarra:2016}


\section{Conclusions}
In this work, we have proposed a workflow for generating configuration-space representations capable of accurately describing single-particle diffusion in complex liquids.
Importantly, the approach does not rely on system-specific or hand-crafted, high-dimensional features for distinguishing solvation states.
Instead, a hidden-Markov-model filtering procedure is applied to generic, low-dimensional solvation features, in order to predict the underlying solvation states.
The resulting filtered trajectories of the solvation features give rise to a structured free-energy landscape, upon which a coarse representation can be easily constructed using standard clustering and kinetic modeling techniques.

The method was applied to a standard model for glassy liquids, where continuous-time random walk models of jumping between local cages are known to characterize the long time diffusion properties.
We demonstrated that the automated configuration-space discretization is capable of quantitatively describing the diffusion constant within the continuous-time random walk framework, without assuming specific characteristics of the jumping process a priori or building in reversal events into the jump definition.
Characterization of the heterogeneity of local diffusion as the system enters the glassy regime motivates the utility of this framework for constructing models that provide insight into the mechanism of diffusion in complex liquids.

More generally, the proposed methodology outlines a data-driven route for detecting the set of relevant many-particle solvation states from low-dimensional order parameters.
In some ways, the philosophy of applying the hidden-Markov-model filter is similar to the inherent structure formalism introduced by Stillinger and Weber---removing thermal noise reveals the relevant minima along the underlying potential energy landscape---but bypasses the expensive determination of these hidden solvation states.
The approach is limited to characterizing activated diffusion processes where the Markovian assumption is valid, although generalizations of the methodology to non-Markovian dynamics may be possible.
Overall, the approach demonstrates significant promise for elucidating mechanisms of heterogenous dynamics in complex liquids.

\vspace{6pt} 

\subsection*{SUPPLEMENTARY MATERIAL}

An online supplement to this article with further methodological details as well as additional results and also an online database consisting of various analysis scripts and input files for the simulations can be found online at \burl{https://github.com/JFRudzinski/Scripts_for_Automated_detection_of_many-particle_solvation_states.git}.

\subsection*{ACKNOWLEDGEMENTS}
We thank Marius Bause, Burkhard D\"unweg, Kurt Kremer, and Roberto Menichetti for critical reading of the manuscript.
We are especially grateful to Kurt Kremer and Roberto Menichetti for useful discussions throughout the development of this work.
This work was funded in part by the TRR 146 Collaborative Research Center of the Deutsche Forschungsgemeinschaft (DFG), the Emmy Noether program supporting TB, a postdoctoral fellowship from the Alexander von Humboldt foundation supporting JFR, and the European Research Council (Grant Agreement No. 340906-MOLPROCOMP) supporting MR.

\bibliography{references_PSU,references_MPIP}


\end{document}